\documentclass[12pt]{article}
\newcommand{\sect}[1]{\section{#1}}

\usepackage{latexsym}
\usepackage{amsfonts}

\def\fraz#1#2{{\strut\displaystyle #1\over\displaystyle #2}}

\def\sca#1#2{({#1}\!\cdot\!{#2})}

\def\ksl{\sca \gamma k}
\def\apsl{\sca\gamma{A'}}

\def\chid{\chi_{{}_{D}}}

\newcommand{\di}[2]{\frac{d {#1}}{d {#2}}}
\newcommand{\de}[2]{\frac{\partial {#1}}{\partial {#2}}}

\def\spazio#1{\vrule height#1em width0em depth#1em}
\def\L{{\cal{L}}}

\date {}
\begin{document}

\title{\bf Spinning Particle with Anomalous Magnetic Moment in an External Plane Wave Field.}

\author{A. Barducci and R.Giachetti}

\maketitle \centerline{{  Department of Physics, University of
Florence and I.N.F.N. Sezione di Firenze }}\centerline{{ Via G.
Sansone 1, I-50019 Sesto Fiorentino, Firenze , Italy\footnote{
e-address: barducci@fi.infn.it, giachetti@fi.infn.it} }}
\bigskip
\bigskip

\centerline{{ Firenze Preprint - DFF - 445/12/07}}

\bigskip
\bigskip
\noindent

\begin{abstract}
In this paper we study the interaction of a Dirac-Pauli particle
with an electromagnetic plane wave, by using a
previously given generalization of the pseudo-classical Lagrangian
for a spinning particle with an anomalous magnetic
moment. We derive the explicit expressions for the eigenfunctions
and the Green's functions of the theory. We discuss the validity
of the semi-classical approach by comparing the wavefunctions with the
(pseudo)-classical solutions of the Hamilton-Jacobi equation.
\end{abstract}
\bigskip
\bigskip

\sect{Introduction.}

\begin{description}
 \item[]
 \end{description}
 Starting with the papers \cite{BM,BDZDVH}, the description of relativistic spinning particles and
 superparticles has found many other different formulations in past years \cite{BDVH}-\cite{GT}.
 The study of these models was firstly originated by their close
 relation with the string theory, but it later became clear
 that the problem was important in itself for a deeper
 understanding of the structure of the quantum theory.
 These models have
 been later investigated and quantized also in the presence of external fields both by
means of canonical quantization and by the use of the
 path-
 integral. More refined results have been achieved when the external fields were taken as plane waves: in these cases it
 has been proved, either by path integral \cite{BBC,BG75,Boud} and by canonical theory \cite{VL,BGJP}, that the complete
 quantum propagator coincides with the results of the semi-classical approximation.

The description of a spinning particle with anomalous magnetic
moment has firstly been given in \cite{B} and later on, with 
different approaches, in \cite{GS1,GS2,BBC2001}. All of these
treatments lead to the same first class Dirac constraints and
therefore to the same quantum wave equation.

The purpose of this work is to study the quantization of a
pseudo-classical spinning particle which has both the normal
magnetic moment and an anomalous magnetic moment in the field
of a plane electromagnetic wave, thus completing the research
begun in \cite{BGJP} for the usual spinning particle. As we shall
show, the main difference with respect to the previous cases is
that now the semi-classical approximation is no longer exact except for
some particular cases: this is due to the interference of the
anomalous magnetic moment with the electric charge, which makes
the presence of a $T$-ordered product in the propagator unavoidable at the quantum level.

We begin in Section 2 with the description of the model for the spinning
particle with anomalous magnetic moment and we derive
the two constraints of the theory and we study the corresponding
canonical quantization procedure. One of these constraints -- the
Dirac-like one --  is linear; the other -- the Klein-Gordon-like
-- is quadratic. Obviously the latter involves some ordering
difficulties in the quantization procedure. However, since the
Poisson bracket of the linear constraint gives rise to the
quadratic constraint, we will quantize the system by following the
Dirac procedure based on the correspondence principle. We recall
that this procedure is based on the postulate that the Poisson
brackets for the the quantum mechanical variables (which are
operators in a Hilbert space) must satisfy the same algebraic
identities as the classical ones; it then follows that the Poisson
brackets are proportional to the (anti)-commutator. We will therefore 
assume this Dirac prescription for quantizing the theory. It
is interesting to notice that the expression we find for the quantized
quadratic constraint, following this procedure, is the same  
we would find by quantizing the theory with the standard Weyl symmetrization
of the quantum dynamical variables. This therefore makes every definition
well posed. In Section 3. we calculate the wavefunction for the
Dirac-Pauli particle and we derive the Green's function
of the theory. Finally, in Section 4, we make a comparison with the semi-classical
approximation that can be obtained by solving the corresponding
pseudo-classical Hamilton-Jacobi equation.

\sect{Pseudoclassical description and canonical quantization.}

As we said in the introduction, there are different approaches to
the description of a pseudo-classical spinning particle both in the
standard case and in the case of a particle with anomalous
magnetic moment. Since all of them lead to the same constraints
and therefore to the same quantum wave equation, we start with the
Lagrangian given in \cite{B}, namely\newpage
\begin{eqnarray}
&{}& \L(x_\mu,\dot x_\mu,\xi_\mu,{\dot \xi}_\mu,\xi_5,{\dot
\xi}_5)=\spazio{1.0}\cr &{}&-\fraz i2 \sca\xi{\dot \xi}-\fraz i2
\xi_5{\dot \xi}_5-q\,\sca{\dot x} A - \Bigl[{m^2
-i(q+\frac{e\mu}2)\,F_{\mu\nu}\xi^\mu\xi^\nu
-\frac{e^2\mu^2}{16m^2}\,F_{\mu\nu}F_{\rho\sigma}\,\xi^\mu\xi^\nu\xi^\rho\xi^\sigma}\Bigr]^{1/2}\spazio{1.0}\cr
&{}& \phantom{XXXXXX} \Bigl[{\Bigl({\dot
x}^\mu-i\,(m+\frac{ie\mu}{4m}\,F_{\mu\nu}\,\xi^\mu\xi^\nu)^{-1}
\,\,\xi^\mu\,({\dot\xi}_5-\frac{e\mu}{2m}\,{\dot
x}_\rho\,F_{\rho\sigma}\,\xi^\sigma)\Bigr)^2\,
}\Bigr]^{1/2} \label{spinlagrangian}
\end{eqnarray}
We use here the standard convention for the metric tensor and for the
gamma matrices \cite{bjorkendrell}.
Here $q$ means the charge of the particle and $e$ the electronic
charge respectively. After using the second class constraints
arising from the Lagrangian, the final form for the two first
class constraints becomes
\begin{equation}
\chid=\sca{(p-qA)}\xi-m\, \xi_5+i\frac{e\mu}{4m}\,F_{\mu\nu}\,\xi^\mu\xi^\nu\xi_5\,,
\label{constr2b}
\end{equation}
\begin{eqnarray}
&{}&\chi=(p-qA)^2-m^2+i(q+\frac {e\mu}2) F_{\mu\nu}\,\xi^\mu\xi^\nu
+i\frac{e\mu}{m}\,(p^\mu-qA^\mu)\,F_{\mu\nu}\,\xi^\nu\xi_5+\spazio{0.8}\cr
&{}&\phantom{XX}\,\frac{e^2\mu^2}{16m^2}\,F_{\mu\nu}F_{\rho\sigma}\,\xi^\mu\xi^\nu\xi^\rho\xi^\sigma,
\label{constr1b}
\end{eqnarray}

The relevant Dirac brackets for the pseudo-classical variables are
\begin{equation}
\{x^\mu,p^\nu\}=-\eta^{\mu\nu}\qquad\{\xi^\mu,\xi^\nu\}=i\eta^{\mu\nu},\qquad\{\xi_5,\xi_5\}=-i,
\end{equation}
leading to the constraint's algebra

\begin{equation}
\{\chid,\chid\}=i\chi,\qquad\{\chid,\chi\}=\{\chi,\chi\}=0
\label{constralg1}
\end{equation}
Because the algebra of the pseudo-classical Poisson brackets is a
graded Lie algebra it follows that the quantum Poisson brackets
must be proportional to an anticommutator for the Fermi-Fermi case
and to a commutator for the Bose-Bose and the mixed Bose-Fermi
cases. The graded quantum commutation rules for the canonical
variables read therefore
\begin{equation}
[x^\mu,p^\nu]=-i\eta^{\mu\nu},\qquad
\{{\hat \xi}^\mu,{\hat \xi}^\nu\}_+=-\eta^{\mu\nu},\qquad\{{\hat \xi}_5,{\hat \xi}_5\}_+=1
\label{constralg2}
\end{equation}
In order to satisfy the anti-commutation relation for the $\hat \xi$
variables, we see that a possible realization can be the
following:
\begin{equation}
{\hat \xi}^\mu=2^{-1/2}\,\gamma_5\gamma^{\mu},\qquad {\hat \xi}_5=2^{-1/2}\,\gamma_5
\label{xigamma}
\end{equation}
The explicit form of the quantized Dirac linear constraint takes
now the form
\begin{equation}
\hat{\chi}_{{}_D}=2^{-1/2}\,\gamma_5\,\Bigl( \sca\gamma{({p}-q{A})}-\frac{e\mu}{8m}\sigma_{\mu\nu}F^{\mu\nu}-m\Bigr)
\equiv2^{-1/2}\,\gamma_5\,\hat{\mathcal{O}}_{{}_D}
\label{constralg}
\end{equation}
where  $\sigma^{\mu\nu}$ is defined by
$\sigma^{\mu\nu}=(i/2)[\gamma^\mu,\gamma^\nu]$.

As we said in the introduction, the quantization of the quadratic
constraint involves some ordering problems, since the constraint
$\chi$ contains terms with product of the momenta and functions of
the coordinates. A correct way of solving this ambiguity is to
define (following the Dirac correspondence principle procedure)
the quantum expression of this constraint according to the
quantized version of equation (\ref{constralg1}), so that
\begin{equation}
\hat{\chi}=-2\hat{\chi}_{{}_D}^2=-\frac12\{\gamma_5\,\hat{\mathcal{O}}_{{}_D},\gamma_5\,\hat{\mathcal{O}}_{{}_D}\}_+ \\
\label{comm_vinc}
\end{equation}
and explicitly
\begin{eqnarray}
 &{}& \hat{\chi}=(p-qA)^2-\frac 12(q+\frac {e\mu}2)\, \sigma_{\mu\nu}F^{\mu\nu}
-i\frac{e\mu}{8m}\sigma^{\mu\nu}\gamma^\rho\partial_\rho F_{\mu\nu}
-\frac{e\mu}{2m}\gamma^\mu\partial^\nu F_{\mu\nu}+\spazio{1.0}\cr
&{}&\phantom{XXXXXXXi}i\frac{e\mu}{2m}\gamma^\mu F_{\mu\nu}(p^\nu-qA^\nu)
-(\frac{e\mu}{8m})^2\sigma_{\mu\nu}\sigma_{\rho\sigma}F^{\mu\nu}F^{\rho\sigma} -m^2
\end{eqnarray}
By now using the relation
$
\sigma^{\mu\nu}\gamma^\rho\partial_\rho F_{\mu\nu}=(i/2)[\gamma^\mu,\gamma^\nu]\,
\gamma^\rho\partial_\rho F_{\mu\nu}=2i\gamma^\mu\partial^\nu F_{\mu\nu}\,,
$
(we recall $\partial^\nu F^* _{\mu\nu}=0$, where $F^* _{\mu\nu}$
is the dual electromagnetic tensor) we will get for the final form
of the quantized quadratic constraint
\begin{eqnarray}
&{}& \hat{\chi}=(p-qA)^2-\frac 12(q+\frac{e\mu}2) F^{\mu\nu}\sigma_{\mu\nu}
-\frac{e\mu}{4m}\gamma^{\mu}\partial^\nu F_{\mu\nu}+\spazio{0.8}\cr
&{}&\phantom{XXXXX}i\frac{e\mu}{2m}\gamma^\mu F_{\mu\nu}(p^\nu-qA^\nu)
-(\frac{e\mu}{8m})^2F^{\mu\nu}F^{\rho\sigma}\sigma_{\mu\nu}\sigma_{\rho\sigma} -m^2
\label{quant_const}
\end{eqnarray}
The commonly adopted approach for solving the previous ordering
ambiguities is to use the Weyl symmetrization prescription
for the classical expressions, which in particular provides the
hermiticity of the corresponding quantum operators. Although in
principle this procedure does not guarantee that the relevant
algebraic structures are maintained at the quantum level, in the
present case (as we said in the introduction) the Weyl
symmetrization procedure would lead to the following quantized
version of the constraint:
\begin{eqnarray}
&{}& \hat{\chi}_{{}_{\rm{Weyl}}}=(p-qA)^2
+i\frac{e\mu}{2m}\{p^\mu-qA^\mu,F_{\mu\nu}\}_+\hat{\xi}^\nu\xi_5
+i(q+\frac{e\mu}2) F^{\mu\nu}\,\hat{\xi}_\mu\hat{\xi}_\nu+\spazio{1.0}\cr
&{}&\phantom{XXXX}(\frac{e\mu}{4m})^2F^{\mu\nu}F^{\rho\sigma}\,\hat{\xi}_\mu\hat{\xi}_\nu\hat{\xi}_\rho\hat{\xi}_\sigma
 -m^2
\label{Weylsym}
\end{eqnarray}
By using the identity
\begin{eqnarray}
&{}&\!\!\!\!\!\!\!\!\!\!\!
\frac12 \{p^\mu-qA^\mu,F_{\mu\nu}\}_+\hat{\xi}^\nu\xi_5=
-\frac14\{p_\mu-qA_\mu,F_{\mu\nu}\}_+\gamma^\nu=
\frac i8 [\sca\gamma{(p-qA)}\,,\,F^{\mu\nu}\sigma_{\mu\nu}]\cr
&{}&\!\!\!\!\!\!\!\!\!\!\!\phantom{\frac12 \{p^\mu-qA^\mu,F_{\mu\nu}\}_+\hat{\xi}^\nu\xi_5}\,
=\frac i4\gamma^\mu(\partial^\nu F_{\mu\nu})+\frac12\gamma^\mu F_{\mu\nu}(p^\nu-qA^\nu)
\end{eqnarray}
it is easily seen that the expression (\ref{Weylsym}) coincides
with (\ref{quant_const}). Therefore we have shown that the
quadratic wave equation appears to be a mere consequence of the
first order wave equation, i.e. the Dirac-Pauli equation.

Finally the equation obtained by squaring the linear Dirac
operator is therefore the following:
\begin{eqnarray}
&{}& \Bigl[(p-qA)^2-\frac 12(q+\frac{e\mu}2) F^{\mu\nu}\sigma_{\mu\nu}
-\frac{e\mu}{4m}\gamma^{\mu}\partial^\nu F_{\mu\nu}+\spazio{1.0}\cr
&{}&\phantom{XXXXXX}i\frac{e\mu}{2m}\gamma^\mu F_{\mu\nu}\,(p^\nu-qA^\nu)
-(\frac{e\mu}{8m})^2\,F^{\mu\nu}F^{\rho\sigma}\sigma_{\mu\nu}\sigma_{\rho\sigma} -m^2\Bigr]\,\psi=0
\end{eqnarray}

We now assume that the electromagnetic potential $A^\mu$ describes
the field of an external plane wave and has thus the following
dependence upon the space-time variables
\begin{eqnarray}
&{}& A^\mu=a^\mu f(\phi)
\label{amu}
\end{eqnarray}
where $\phi=\sca kx$ with $k^2=0$ and where we assume the Lorentz
gauge, so that also $\sca ka=0$. From this potential we get the
electromagnetic tensor ($f'(\phi)=df(\phi)/d\phi$)
\begin{eqnarray}
&{}& F^{\mu\nu}=f^{\mu\nu} f'(\phi)\,,~~~f^{\mu\nu}=k^\mu a^\nu-k^\nu a^\mu
\end{eqnarray}
and the final expression for the wave equation to be solved
reads
\begin{eqnarray}
&{}& \!\!\!\!\!\!\!\!\!\! \Bigl[\partial^2 +2iq\sca A\partial-q^2A^2+i(q+\frac{e\mu}2)\ksl \apsl
\spazio{1.0}\cr
&{}& \!\!\!\!\!\!\!\!\!\!\phantom{XXXX}- i\frac{e\mu}{2m}
\Bigl(i\ksl \sca{{A}'}\partial
-q\ksl \sca{{A}'}{{A}}
-i\apsl \sca k\partial\Bigr)+m^2\Bigr]\psi(x)=0
\end{eqnarray}

\sect{The wave function and the Green's function} Recalling the
standard derivation of the Volkov solutions for the usual Dirac
and Klein-Gordon equations \cite{VL,kibble}, we make for the wave
function the ansatz
\begin{eqnarray}
\psi(x)=\frac 1{(2p_0V)^{1/2}}\exp\Bigl\{-i\sca px-\frac
i{2\sca kp}\int\limits_{-\infty}^{ (k\cdot x)} d\phi\Bigl[2q\sca
Ap-q^2A^2\Bigr]\Bigr\}\,M(\sca kx)u(p,s) \label{ansatz}
\end{eqnarray}
where $M(\sca kx)$ is  a $4\times 4$ unknown Dirac matrix and where the initial condition
\begin{eqnarray}
&{}& \psi(x)\matrix{{}\cr \longrightarrow\cr{}^{x_0\rightarrow-\infty}} (2p_0V)^{-1/2}\,\exp\Bigl[-i\sca px\Bigr]\,u(p,s)
\label{cond_iniz}
\end{eqnarray}
is assumed. In equation (\ref{cond_iniz}) $u(p,s)$ is a constant
spinor which is is a solution of the corresponding free Dirac
equation. Moreover, since the redefinition of $p^\mu$ as
$p^\mu+\alpha k^\mu$ leaves invariant the equation (\ref{ansatz})
due to the fact that  $M(\sca kx)$ is a unknown function to be determined, we can
impose on the constant vector $p^\mu$ the mass-shell relation
$p^2=m^2$. The differential equation satisfied by $M$ is therefore
\begin{eqnarray}
&{}& \!\!\!\!\!\!\!\!\!\frac{dM}{d\phi}=\frac 1{2\sca k p}\Bigl[(q+\frac{e\mu}2){\ksl}\apsl\spazio{1.0}\cr
&{}&\!\!\!\!\!\!\!\!\!\qquad\qquad\qquad\qquad -\frac{e\mu}{2m}\,\Bigl(\sca {A'}p{\ksl}-
q\sca {A'}A{\ksl}-\sca kp\apsl\Bigr)\Bigr]\,M(\phi)
\label{dMdphi}
\end{eqnarray}
and the formal solution of the previous equation is
\begin{eqnarray}
&{}& \!\!\!\!\!\!\!\!\!{M}(\sca kx)=T\Bigl\{\exp\Bigl[\frac 1{2\sca k p}
\int\limits_{-\infty}^{(k\cdot x)}\,d\phi\,
\Bigl((q+\frac{e\mu}2){\ksl}\apsl - \spazio{1.0}\cr
&{}& \!\!\!\!\!\!\!\!\!\phantom{XXXXXXXXXXX}\frac{e\mu}{2m}\,\Bigl(\sca{A'}p{\ksl}-
q\,\sca{A'}A{\ksl}-\sca kp\apsl\Bigr)\Bigr)\,\Bigr]\Bigr\}
\label{Mrisolta}
\end{eqnarray}
with the asymptotic condition
$
 M(-\infty)={\mathbf{1}}\,.
$
The final expression for the wave equation reads
\begin{eqnarray}
 &{}&\!\!\!\!\!\!\!\!\!\!\psi(x)=\frac 1{(2p_0V)^{1/2}}\,\exp\Bigl[
-i\sca px-\frac i{2\sca kp}\int\limits_{-\infty}^{k\cdot x}\,d\phi\,\Bigl(2q\,\sca Ap-q^2A^2\Bigr)\Bigr]\,
  \cr
&{}&~~~~~
T\Bigl\{\exp\Bigl[\,\frac1{2\sca kp}\int\limits_{-\infty}^{(k\cdot x)}d\phi\,
\Bigl((q+\frac{e\mu}2){\ksl}\apsl-\frac{e\mu}{2m}\,(\sca{A'}p{\ksl}-\spazio{1.4}\cr
&{}& \qquad\qquad\qquad\qquad\qquad\qquad q\,\sca{A'}A{\ksl}-\sca kp\apsl)\Bigr)\,\Bigr]\Bigr\}\,u(p,s)
\label{soluz_psi}
\end{eqnarray}

The expression (\ref{soluz_psi}) for $\psi(x)$ is obviously a formal
one. However there are at least two special cases in which from
(\ref{soluz_psi}) we can get a closed form for this solution.
The first of these cases occurs when $q\neq 0$ and  $\mu=0$, that
is the special case of no anomalous magnetic moment but a charged
particle. This choice leads immediately to the Volkov solution
\begin{eqnarray}
\psi(x)=\frac 1{(2p_0V)^{1/2}}\,\exp\Bigl\{
-i\sca px-\frac i{2\sca kp}\,\int\limits_{-\infty}^{(k\cdot x)}\,d\phi\,\Bigl(2q\,\sca Ap -q^2A^2+\frac q2
\,F^{\mu\nu}\sigma_{\mu\nu}\Bigr)
\Bigr\} \,u(p,s)
\end{eqnarray}

The second case, with $q= 0$ and $\mu\neq0$, i.e. the special case
of pure anomalous magnetic moment and no charge, gives a new
result in a closed form. Indeed we obtain
\begin{eqnarray}
 &{}&\!\!\!\!\!\!\!\!\!\!\!\!\!\!\!\!\!\!\!\!\psi(x)=\frac 1{(2p_0V)^{1/2}}\,\exp\Bigl\{
-i\sca px+\frac{e\mu}{4\sca kp}\int\limits_{-\infty}^{(k\cdot x)}d\phi\,
\Bigl[{\ksl}\sca\gamma{A'}-\spazio{1.0}\cr
&{}&\phantom{XXXXXXXXXXXX}\frac 1m\Bigl(
\sca{A'}p{\ksl}
-\sca{k}{p}\sca\gamma{A'}\Bigr)
\Bigr]\Bigr\} \,u(p,s)
\end{eqnarray}
It can be realized that the presence of the time ordering is due to the joint action of the
terms involving the anomalous magnetic moment $\mu$ and the electric charge $q$. When only one of these
components is effective,
the time ordering can be calculated and an expression in closed form is obtained.

Similar procedures to the ones we have used for calculating the
wavefunction, can be used also for the computation of the Green's
function $S_{{}_F}(x,y)$ that satisfies
\begin{eqnarray}
{\hat {\mathcal O}}_{{}_D}\,S_{{}_F}(x,y)=\delta^4(x-y)
\label{sf_eq}
\end{eqnarray}
where ${\hat {\mathcal O}}_{{}_D}$ is given in (\ref{constralg}).
As usual, instead of solving directly  the first-order Dirac
equation (\ref{sf_eq}), it is easier to consider the corresponding
second order equation
\begin{eqnarray}
{\hat \chi}\,\Delta_{{}_F}(x,y)=\delta^4(x-y)
\end{eqnarray}
where $\Delta_{{}_F}(x,y)$ is defined by (see eq.(\ref{comm_vinc}))
\begin{eqnarray}
\gamma_5S_{{}_F}(x,y)\gamma_5=-{\hat {\mathcal O}}_{{}_D}\Delta_{{}_F}(x,y)
\label{sfdf}
\end{eqnarray}
The general second-order wave equation for the Green's function is
therefore
\begin{eqnarray}
 &{}&\!\!\!\!\!\!\!\!\!\!\!\!\!\!\!\!\!\!\!\! \Bigl[\partial^2+2iq\,\sca A\partial - q^2A^2+\frac 12(q+\frac{e\mu}{2})\,F^{\mu\nu}\sigma_{\mu\nu}+\frac{e\mu}{4m}\gamma^\mu\partial^\nu F_{\mu\nu}
-\spazio{1.0}\cr
&{}& 
\frac{i\mu e}{2m}\gamma^\mu F_{\mu\nu}(i\partial^\nu-qA^\nu)+
\Bigl(\frac{e\mu}{8m}\Bigr)^2F^{\mu\nu}F^{\rho\tau}
\sigma_{\mu\nu}\sigma_{\rho\tau}+m^2\Bigr]\,\Delta_{{}_F}(x,y)=-\delta^4(x-y)
\label{green_eq_1}
\end{eqnarray}
and, by choosing $A^\mu$ as in (\ref{amu}) we obtain,
\begin{eqnarray}
 &{}& \!\!\!\!\!\!\!\!\!\!\!\!\!\!\!\!\!\!\Bigl[\partial^2+2iq\,\sca A\partial - q^2A^2+i(q+\frac{e\mu}{2}){\ksl}\apsl-\spazio{1.0}\cr
&{}&\!\!\!\!\!\!\!\!\!\! i\frac{e\mu}{2m}\Bigl(i{\ksl}\sca {A'}\partial -i\apsl \sca k\partial -q{\ksl}\sca{A'}A\Bigr)
+m^2\Bigr]\,\Delta_{{}_F}(x,y)=-\delta^4(x-y)
\label{green_eq_2}
\end{eqnarray}

To solve (\ref{green_eq_2}) we assume as ansatz
\begin{eqnarray}
 \Delta_{{}_F}(x,y)=\int\,\frac{d^4p}{(2\pi)^4}\,\frac{\exp\{-i\sca p{(x-y)}\}}{p^2-m^2}\,\exp\{-iF_{{}_1}(\phi_x,\phi_y)\}\,
N(\phi_x,\phi_y)
\label{ansatz_DeltaF}
\end{eqnarray}
where $\phi_x=\sca kx$, $\phi_y=\sca ky$ and
\begin{eqnarray}
 F_{{}_1}(\phi_x,\phi_y)=\frac 1{2\sca kp}\int\limits_{(k\cdot y)}^{(k\cdot x)}\,d\phi\,\Bigl(2q\,\sca Ap - q^2A^2\Bigr)
\label{f1}
\end{eqnarray}
By inserting the ansatz (\ref{ansatz_DeltaF}) into (\ref{green_eq_2}) we obtain the following equation for $N(\phi_x,\phi_y)$:
\begin{eqnarray}
&{}&\!\!\!\!\!\!\!\!\!\!\!
\int\,\frac{d^4p}{(2\pi)^4}\,\exp\{-i\sca p{(x-y)}\}\Bigl[
1-\exp\{-iF_{{}_1}(\phi_x,\phi_y)\}\,N(\phi_x,\phi_y)
\Bigr] =\spazio{1.0}\cr
&{}&\!\!\!\!\!\!\!\!\int\frac{d^4p}{(2\pi)^4}\exp\{-i\sca p{(x-y)}-iF_{{}_1}(\phi_x,\phi_y)\}\Bigl(\frac{-2i\sca kp}{p^2-m^2}\Bigr)
\Bigl[\frac{dN}{d\phi_x}-\frac 1{2\sca kp}\Bigl(
(q+\frac{e\mu}{2}){\ksl}\apsl-
\spazio{1.6}\cr
&{}&\phantom{XXXXX}
\frac{e\mu}{2m}\Bigl({\ksl}\sca{A'}p -q{\ksl}\sca{A'}A-\apsl \sca kp\Bigr)
\Bigr)\,N(\phi_x,\phi_y)
\Bigr]
\label{equa_int_N}
\end{eqnarray}
We can now observe that the expression in the square bracket in the
right hand side of equation (\ref{equa_int_N}) coincides with
(\ref{dMdphi}), up to the obvious substitutions. The right hand
side of (\ref{equa_int_N}) is therefore vanishing by assuming for 
$N(\phi_x,\phi_y)$ the form
\begin{eqnarray}
&{}&\!\!\!\!\!\!\!\!\!\!\!\!N(\phi_x,\phi_y)=T\Bigl\{
\exp\Bigl[
\frac 1{2\sca kp}\int\limits_{\phi_y}^{\phi_x}
(q+\frac{e\mu}{2}){\ksl}\apsl-\spazio{1.0}\cr
&{}&\phantom{XXXXXXXXXXXX}\frac{e\mu}{2m}\Bigl({\ksl}\sca{A'}p -q{\ksl}\sca{A'}A-\apsl \sca kp\Bigr)
\Bigr]
\Bigr\}
\end{eqnarray}
We have therefore to verify that the left hand side of
(\ref{equa_int_N}) is also vanishing.
 We first separate in the left hand side of
(\ref{equa_int_N}) the component of $p_\mu$ in the direction of
$k_\mu$,
$~p^\mu = {p\,'\,}^\mu + \alpha\, k^\mu \label{pandk}~$
where $p'^{\,\mu}$ spans a three-dimensional surface, \cite{kibble}. Since the
function $F_{{}_1}(\phi_x,\phi_y)$ is independent of $\alpha$ we
can integrate over $\alpha$, getting
\begin{eqnarray}
&{}& \!\!\!\!\!\!\!\!\!\!\int{d\alpha\over 2\pi} \,{\exp\Bigl[{-i \alpha (\phi_x-\phi_y)}\Bigr]
\Bigl(\, 1- \exp\Big[{-iF_{{}_1}(\phi_x,\phi_y)}}\Bigr]\,\Bigr) =\cr
&{}&\phantom{XXXXXXXXXXX}\delta(\phi_x-\phi_y)\,\Bigl(\,1-
{\exp\Big[{-iF_{{}_1}(\phi_x,\phi_y)}}\Bigr]\,\Bigr)\,.
\label{verificaf1}
\end{eqnarray}
Due to (\ref{f1}), the previous expression is clearly vanishing and
thus the left hand side of (\ref{equa_int_N}) is  vanishing too.
The Green's function $\Delta_{{}_F}(x,y)$ has therefore the final
form
\begin{eqnarray}
&{}& \Delta_{{}_F}(x,y)=\int\,\frac{d^4p}{(2\pi)^4}\,\frac{1}{p^2-m^2}\,
\exp\Bigl[-i\sca p{(x-y)}-\frac i{2\sca kp}\int\limits_{(k\cdot y)}^{(k\cdot x)}\,d\phi\,\Bigl(2q\,\sca Ap - q^2A^2
\Bigr)\Bigr]\cr
&{}& \phantom{XXXX}T\Bigl\{
\exp\Bigl[
\frac 1{2\sca kp}\int\limits_{(k\cdot y)}^{(k\cdot x)}d\phi\,
(q+\frac{e\mu}{2}){\ksl}\apsl-\cr
&{}&  \phantom{XXXXXXXXXXXXXX}\frac{e\mu}{2m}\Bigl({\ksl}\sca{A'}p -q{\ksl}\sca{A'}A-\apsl \sca kp\Bigr)
\Bigr]
\Bigr\}
\end{eqnarray}
and $S_{{}_F}(x,y)$ can be immediately obtained by (\ref{sfdf}).
The discussion we have done on the special cases $q\neq 0\,,~\mu=0$ and $q=0\,,~\mu\neq 0$ when solving for the wavefunction $\psi(x)$, holds for the Green's function too.
\bigskip

\sect{The Hamilton-Jacobi equation}

In this section we want to comment on some aspects of the
pseudo-classical nature of the solutions we have determined. The
discussion involves in a natural way the properties of the
Hamilton-Jacobi equation and the corresponding characteristic
functions. We recall that, according to Dirac, the extended
Hamiltonian is written as a linear combination of the first-class
constraints ${\chi}$ and ${\chi}_{{}_D}$ with arbitrary
coefficients, namely
\begin{equation}
 {\cal H}=\alpha_{{}_D}\chid+\alpha_{{}_1}\chi
\end{equation}
where $\alpha$ and $\alpha_{{}_1}$ are arbitrary Lagrangian multipliers.
In particular, the choice $\alpha_{{}_D}=0$ and $\alpha_{{}_1}\neq 0$ yields
\begin{eqnarray}
 &{}&{\cal H}=\alpha_{{}_1}\Bigl[(p-qA)^2+\frac{ie\mu}m(p_\mu-qA_\mu)F^{\mu\nu}\xi_\nu\xi_5-m^2+
i(q+\frac{e\mu}2)F^{\mu\nu}\xi_\mu\xi_\nu+\spazio{1.0}\cr
 &{}&\phantom{XXXXXXXXXXXX}\frac{\mu^2 e^2}{16m^2}F^{\mu\nu}F^{\rho\sigma}\xi_\mu\xi_\nu\xi_\rho\xi_\sigma
\Bigr]
\end{eqnarray}
and the corresponding covariant Hamilton-Jacobi equation turns out to be
\begin{eqnarray}
&{}& \!\!\!\!\!\!\!\!\!\!\!\!\!\Bigl(\de{S}{x^\mu}+qA_\mu\Bigr)^2-\frac{i\mu e}{m}\,\Bigl(\de{S}{x^\mu}+qA_\mu\Bigr)(k^\mu a^\nu-k^\nu a^\mu)\,\xi_\nu\xi_5\,f'-m^2+\spazio{1.2}\cr
&{}& \qquad\qquad\qquad\qquad\qquad\qquad\qquad\quad 2i(q+\frac{e\mu}{2})\sca k\xi\sca a\xi\,f'=0
\label{HJE}
\end{eqnarray}

The structure of the Dirac extended Hamiltonian suggests, for the pseudo-classical characteristic function, the factorized form
\begin{eqnarray}
S=-\sca px-\frac 1{2\sca kp}\int\limits_{-\infty}^{(k\cdot x)}d\phi\,\Bigl(2q\,\sca Ap -q^2A^2\Bigr) + S_{{}_D}(\sca kx)
\end{eqnarray}
as a function of the coordinates $x^\mu$ and of the initial
momentum $p^\mu$ and $p^\mu$ satisfies the mass-shell condition $p^2=m^2$. 
We therefore obtain for the unknown term $
S_{{}_D}(\sca kx)$ the following first-order differential equation
\begin{eqnarray}
&{}&\di{S_{{}_D}}{\phi}+\frac{i\mu e}{2m\sca kp}
\Bigl[-\sca kp\sca{A'}\xi\xi_5+\sca{A'}p\sca k\xi\xi_5-q\sca{A'}A\sca k\xi\xi_5\Bigr]
\spazio{1.0}\cr
&{}&\qquad\qquad\qquad\qquad  -\frac{i}{\sca kp}(q+\frac{e\mu}{2})\sca k\xi\sca{A'}\xi=0
\end{eqnarray}
The complete solution for the characteristic function of the Hamilton-Jacobi equation is then
\begin{eqnarray}
&{}& \!\!\!\!\!\!\!\!\!\!\!\!\!\!\!\! S=-\sca px-\frac 1{2\sca kp}\int\limits_{-\infty}^{(k\cdot x)}d\phi \Bigl(2q\,\sca Ap -q^2A^2\Bigr)+
\frac{i}{\sca kp}\int\limits_{-\infty}^{(k\cdot x)}d\phi\Bigl[(q+\frac{e\mu}{2})\sca k\xi \sca{A'}\xi-\spazio{1.0}\cr
&{}& \qquad\qquad\qquad\qquad\qquad  \frac{e\mu}{2m}\Bigl(-\sca kp\sca{A'}\xi\xi_5+\sca{A'}p\sca k\xi\xi_5
-q\sca{A'}A\sca k\xi\xi_5\Bigr)\Bigr]
\end{eqnarray}
Taking into account the realization (\ref{xigamma}) for the
quantized Grassmann variables we finally obtain for the action of
 $\exp(i\hat S)$ on the constant spinor $u(p,s)$
\begin{eqnarray}
&{}& \!\!\!\!\!\!\!\!\!\!\!
\frac{\exp(i{\hat S})}{(2p_0V)^{1/2}}\,\,u(p,s)=\frac{1}{(2p_0V)^{1/2}}\,
\exp\Bigl[
-i\sca px-\frac i{2\sca kp}\int\limits_{-\infty}^{(k\cdot x)}d\phi \Bigl(2q\,\sca Ap-q^2A^2\Bigr)\Bigr]\spazio{1.0}\cr
&{}&  \qquad
\exp\Bigl[
\Bigl(\frac{1}{2\sca kp}\int\limits_{-\infty}^{(k\cdot x)}d\phi
\Bigl[(q+\frac{e\mu}{2}){\ksl}\apsl-\spazio{1.0}\cr
&{}&\qquad\qquad\qquad\qquad\quad \frac{e\mu}{2m}\Bigl(-\sca kp\apsl+\sca{A'}p{\ksl}-q\sca{A'}A{\ksl}\Bigr)\Bigr]\,u(p,s)
\label{espis}
\end{eqnarray}
 where $\hat S$ is the quantized version of $S$.
 
Contrary to the results previously known in literature for the
scalar and Dirac particles \cite{BG75,VL,BGJP}, we can see
that the semi-classical results are not now coincident with the
general quantum solutions, but for the two particular cases we
have previously discussed.



\end{document}